\documentclass[seceq]{ptptex}
\usepackage{graphicx,color}

\title{
The CKM Matrix from Lattice QCD%
}

\author{
Paul B. \textsc{Mackenzie}%
}

\inst{
Fermilab, Batavia, IL  60510, USA
}

\abst{
Lattice QCD plays an essential role in testing and determining the parameters of
 the CKM theory of flavor mixing and CP violation.
Very high precisions are required for lattice calculations
analysing CKM data; I discuss  the prospects for achieving them.
Lattice calculations will also play a role in investigating
flavor mixing and CP violation beyond the Standard Model.
}

\begin{document}

\maketitle

\section{Introduction}

The Cabibbo-Kobayashi-Maskawa (CKM) matrix\cite{Kobayashi:1973fv,Cabibbo:1963yz}
 parametrizes the couplings between
flavors of quarks under the weak interactions.
Quarks are believed to be permanently confined within hadrons,
and the dynamics of quarks and gluons in hadrons is nonperturbative.
Lattice quantum chromodynamics (QCD)
 is the only general method for performing nonperturbative
calculations in QCD, so it is not surprising that many analyses of 
the physics of the CKM matrix require lattice QCD calculations.

If the Standard Model of particle physics is solely responsible for flavor mixing,
the CKM matrix will be unitary, and parametrizable with four parameters, 
such as those of the Wolfenstein parametrization,\cite{Wolfenstein:1983yz}
 $A$, $\lambda$, $\overline{\rho}$, and $\overline{\eta}$:
\begin{equation}
\label{eq:CKM}
V=
\left(
\begin{array}{ccc}
V_{ud}&V_{us}&V_{ub}\\
V_{cd}&V_{cs}&V_{cb}\\
V_{td}&V_{ts}&V_{tb}
\end{array}
\right)
\approx
\left(
\begin{array}{ccc}
1-\frac{1}{2}\lambda^2  & \lambda & A\lambda^3(\overline{\rho}-i\overline{\eta})\\
 -\lambda & 1-\frac{1}{2}\lambda^2  & A\lambda^2 \\
 A\lambda^3(1-\overline{\rho}-i\overline{\eta}) & -A\lambda^2 &  1
\end{array}
\right)
\end{equation}

\begin{figure}
       \centerline{\includegraphics[width=.65\textwidth]{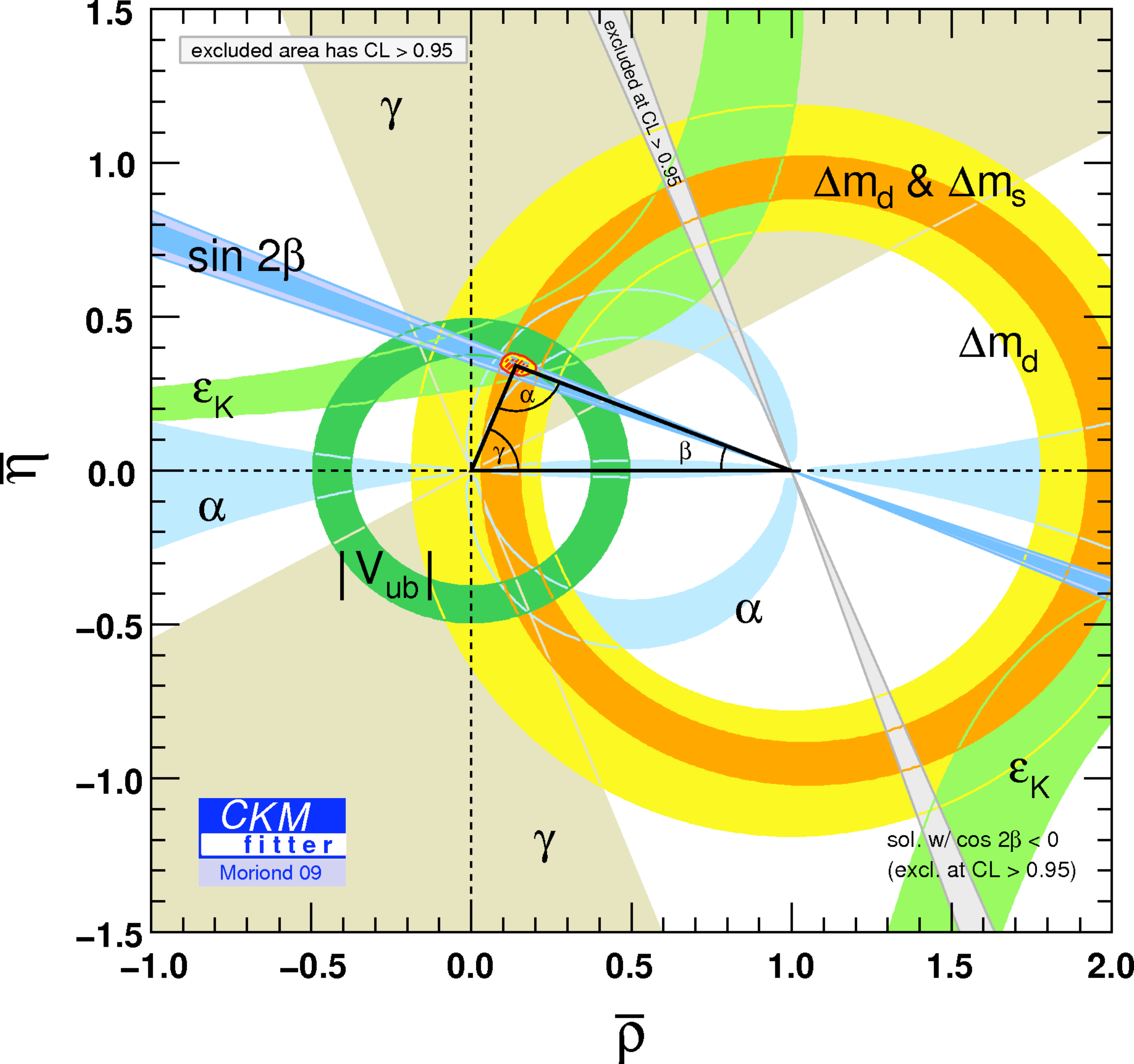}}
\caption{Bounds in the $\overline{\rho}$-$\overline{\eta}$ plane from a global
fit.  (CKMfitter\cite{ckmfitter}.)
Lattice QCD is responsible for the bounds from
$K\overline{K}$ mixing (light green band)
and from $B\overline{B}$ and $B_s\overline{B}_s$ mixing
(yellow and orange bands).
It helps with the bound from $|V_{ub}|$ (dark green band).
}
\label{fig:rhoeta}
\end{figure}

The rows and columns of unitary matrices are orthogonal, satisfying
\begin{equation}
\sum_{k}V_{ik}V^*_{jk} = \sum_{k}V_{ki}V^*_{kj} = \delta_{ij}
\end{equation}
The off-diagonal combinations vanish, and can be represented by triangles
in the complex plane.
The best-known of these is
\begin{eqnarray}
 V_{ud}V^*_{ub}+V_{cd}V^*_{cb}+V_{td}V^*_{tb} &=&0.
\end{eqnarray}
If this equation is divided by $V_{cd}V^*_{cb}$ and parametrized as in
the right side of
Eq.~(\ref{eq:CKM}), one obtains the familiar
$\overline{\rho}$-$\overline{\eta}$ unitarity triangle of Fig.~\ref{fig:rhoeta}.
If the CKM matrix is the full story of flavor physics and the matrix
is unitary, measurements of $\overline{\rho}$ and $\overline{\eta}$ 
should all be consistent,  whether they are
obtained via $V_{ub}$, $V_{td}$, or through some other constraints.
Conflicting determinations of $\overline{\rho}$ and $\overline{\eta}$ 
would constitute evidence for physics beyond the Standard Model.

Lattice analyses of meson decays and mixings provide some 
of the most important ways of determining CKM matrix elements.
Fig.~\ref{fig:rhoeta} shows some of the most important constraints on
$\overline{\rho}$ and $\overline{\eta}$.
In $K\overline{K}$  mixing  (light green band), and in
$B\overline{B}$ and $B_s\overline{B}_s$ mixing (yellow and orange bands),
lattice QCD is the only first-principles way of determining the nonperturbative
parameters that relate the quark scattering amplitudes of the Standard Model
Lagrangian to the meson mixing parameters observed by experiment.
In meson decays, determinations of CKM elements from
 nonperturbative lattice calculations of leptonic and
exclusive semileptonic decay amplitudes provide an important complement to 
determination from inclusive decays via perturbative QCD. 
All of the CKM matrix elements except $V_{tb}$ can be determined from one
of these exclusive processes with lattice QCD. 
Table~\ref{fig:CKM} shows the CKM matrix elements and
the process to which each is most sensitive: meson leptonic decay constants, $f_M$,
exclusive semileptonic decays, $M_1\to M_2 l \nu$, and meson-antimeson mixing amplitudes,
$<M|\overline{M}>$.
\begin{table}[b]
\begin{center}
$
\color{blue}
\left(
\begin{array}{ccc}
V_{ud}&V_{us}&V_{ub}\\
\color{green} f_\pi  &\color{green}   f_K  & \color{green} f_B\\
       &\color{green} K\rightarrow\pi l \nu& \color{green} B\rightarrow\pi l \nu  \\
V_{cd}&V_{cs}&V_{cb}\\
\color{green}f_D  & \color{green}f_{D_s}  & \color{green} B\rightarrow D^* l \nu   \\
 \color{green}D\rightarrow \pi l \nu & \color{green}D\rightarrow K l \nu &\color{green} B\rightarrow D l \nu  \\
V_{td}&V_{ts}&\color{black}V_{tb}\\
\color{green}<B|\overline{B}>  & \color{green}<B_s|\overline{B_s}> &  \\ 
\end{array}
\right)
$
\end{center}
\label{fig:CKM}
\caption{All of the CKM matrix elements except $V_{tb}$ can be determined
 from a tractable lattice QCD calculation.}
\end{table}

Lattice Quantum Chromodynamics is a tool for understanding strongly
interacting QCD at low energies, where QCD perturbation theory fails.
It is an essential tool for studying weak interactions in low-energy
hadronic physics because the effects of weak interactions involving quarks
are shrouded at low energies by the effects of the strong interactions.
These strong-interaction effects must be understood quantitatively before
the weak interaction properties can be inferred.
For example, when a meson decays weakly into leptons, the decay amplitude
is proportional to both a CKM matrix element and a hadronic decay constant,
which parametrizes the amplitude for the two valence quarks in a meson to 
interact at a point.

In lattice QCD, quantum fields for quarks and gluons are defined on the
sites and links of a four-dimensional space-time lattice.
The quantum fluctuations of the fields described by the field theory path 
integral are calculated with Monte Carlo methods.
The physical theory is defined as the zero-lattice-spacing limit of the
lattice theory.  Because the number of degrees of freedom in the path integral
becomes infinite in this limit, this limit is computationally expensive.

In the last decade, due to improvements in methods, in algorithms, and in
computers,
 lattice QCD calculations have become able to produce
serious, first-principles results for many simple but important quantities.
Prime among these for CKM physics are the decay constants,
exclusive semileptonic decays, and meson-antimeson mixing amplitudes 
of stable mesons.
Stable pseudoscalar mesons are among the most tractable quantities for 
current lattice methods.  They have the smallest and best controlled
uncertainties for statistical errors, finite volume errors, and other
quantities.
Since they also provide some of the most accurate CKM-related experimental
results, they provide the most accurate determinations of the CKM matrix
from lattice QCD.

Current lattice calculations are done on computers that are a factor
of $10^{8}$ more powerful than the VAX 11/780s on which the first
numerical lattice QCD calculations were done.
Remarkably,  just as large a factor of improvement has come from improvements
in algorithms, and an even larger factor has come from improved methods.
This has made it possible to abandon the tactic in early lattice calculations
of ignoring the effects of quark-antiquark pairs (the ``quenched'' approximation).
The most serious lattice calculations are now all unquenched.
Lattice calculations contain errors due to discretization, to extrapolation to
the chiral limit, to operator normalizations, etc.
Estimating the expected precision in light of these uncertainties  
is a key element in serious lattice calculations.
The errors due to discretization have been greatly ameliorated by the use of
improved actions.
There are a half dozen families of actions for discretizing fermion fields
with widely disparate virtues and drawbacks.
Which, if any, is best for any given purpose is still contentious.
The good news for observers and customers of lattice calculations is that many
important quantities are becoming available in several types of quark methods,
so that outsiders can see for themselves how well they agree.

Most quantities that have been calculated so far with lattice QCD
agree well with nonlattice results to the expected precision.
A few, however, do not, and these are some of the more interesting cases
in current lattice phenomenology.
For example, lattice calculations of $f_{D_s}$, the 
leptonic decay constant of the $D_s$ meson, lie significantly
 below the experimental results,  while
lattice calculations of other decay constants are spot on.
Determinations of CKM matrix elements from inclusive $B$ decays differ by
around two sigma from those with lattice calculations of the 
exclusive semileptonic decays
$B\rightarrow D^* l \nu$ and $B\rightarrow \pi l \nu$.
I will describe these results as well as the much larger body of results
for which everything is very consistent.

\section{Lattice Determinations of the CKM Matrix Elements
}
Flavor physics was reviewed at Lattice 2008 by
 Lellouch\cite{Lellouch:2009fg} (for kaons) and by Gamiz\cite{Gamiz:2008iv}
(for heavy flavor).
I will often refer to their talks for the state of lattice CKM data,
and will often use their averages rather than provide a complete review here.
I will discuss only unquenched lattice results, mostly with three light flavors,
sometimes with only two.
I will discuss all of the CKM matrix elements, but I will go into more detail
where there is something interesting to say,
and discuss others more briefly.

In this article, I am grouping together CKM matrix elements  which 
can be derived from related physical quantities
and calculated with related methods.
For each set of elements, I will then discuss methods to determine them.
In the next four subsections, I discuss:
\begin{itemize}
\item $V_{cd}$ and $V_{cs}$.  These may be obtained from
lattice calculations of leptonic and
exclusive semileptonic decays of $D$ and $D_s$ mesons.
\item $V_{ub}$ and $V_{cb}$.  These may be obtained from leptonic and
exclusive semileptonic decays of $B$ mesons.
\item $V_{ud}$ and $V_{us}$.  These are obtained from pion and kaon 
leptonic and semileptonic decays.
\item $V_{td}$ and $V_{ts}$.  These are obtained from the meson-antimeson
mixings $<B|\overline{B}>$ and $<B_s|\overline{B}_s>$.  I also discuss 
$<K|\overline{K}>$ mixing in this section.
\end{itemize}
These quantities constitute the core of lattice CKM phenomenology.

\subsection{$V_{cd}$ and $V_{cs}$
}

$|V_{cd}|$ and $|V_{cs}|$ can be derived from leptonic and
exclusive semileptonic decays of $D$ and $D_s$ mesons.
In both cases, the processes are related to each other 
 by SU(3) flavor symmetry.

\subsubsection{$f_D$, $f_{D_s}$}
$f_{D_s}$ presents one of the very few disagreements between lattice 
calculations and other results.
I will therefore discuss these calculations in detail.
The HPQCD collaboration has calculated the four decay constants
$f_\pi$, $f_K$, $f_D$, and $f_{D_s}$\cite{Follana:2007uv}
 with an
improved form of staggered fermions called ``HISQ'' quarks (``highly improved
staggered quarks'').\cite{Follana:2006rc}
Staggered fermions have multiple poles in quark propagators which cause a 
doubling of quark flavors, and whose effects must be removed from
physical calculations.
Transitions between these multiple poles (or ``tastes'' in lattice jargon)
cause significant discretization errors.
The usual improved staggered fermions (so-called ``asqtad'' fermions)
remove these taste breaking effects at the
one-gluon level.  
HISQ quarks remove two-gluon scattering effects as well.
In addition, to apply HISQ to charm quarks, HPQCD has calculated the 
${\cal O}(ma^p)$ errors to a high order and removed them.
($(ma)^4$ turned out to suffice for sub-per cent precision.)
These improvements allowed the calculation of the four decay constants to higher
precision than ever before.
They obtained:\cite{Follana:2007uv}
\begin{eqnarray}
  f_\pi &=& 157(2) {\rm MeV} \\
  f_\pi/f_K &=& 1.189(7)      \\
  f_D &=& 207(4){\rm MeV}     \\
  f_{D_s} &=& 241(3){\rm MeV}. 
\end{eqnarray}
The first three agree with experiment to within the stated precisions, a few
per cent,
but $f_{D_s}$ presents a puzzle, as I now discuss.

\begin{figure}
       \centerline{\includegraphics[width=\textwidth]{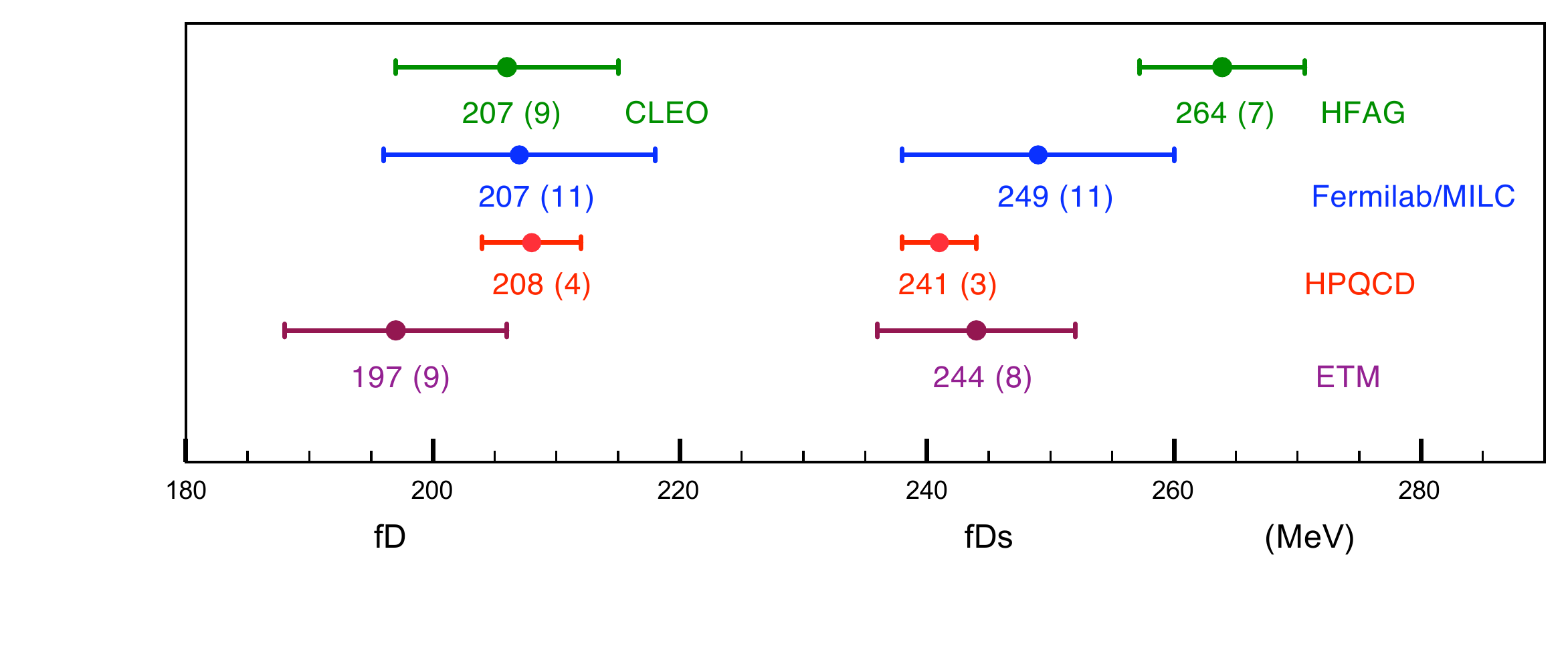}}
\caption{Theory and experiment for $f_D$ and $f_{D_S}$.
Theory results from HPQCD\cite{Follana:2007uv}, 
Fermilab/MILC\cite{Bernard:2009wr,Bernard:2007zz}, 
and ETM\cite{Blossier:2009bx}.
Experimental results from CLEO\cite{:2008sq} ($f_D$) 
and HFAG\cite{hfagDs} ($f_{D_s}$).}
\label{fig:fD(s)}
\end{figure}

$f_\pi$, $f_K$, $f_D$, and $f_{D_s}$ have also been calculated
by the European Twisted Mass Collaboration using twisted-mass fermions.
 Fermilab/MILC have calculated $f_D$ and $f_{D_s}$ using asqtad light quarks.
These groups obtain results compatible with HPQCD,  but with less precision.
ETM obtain\cite{Blossier:2009bx}
 $f_{D_s}= 244(8)$ MeV and $f_D=197(9)$ MeV.
(I will return to the subject of pion and kaon decay constants in a later section.)
Fermilab/MILC, using clover/Fermilab charm quarks and asqtad light 
quarks obtain\cite{Bernard:2009wr,Bernard:2007zz}
$f_D=207(11)$ MeV and $f_{D_s}=249(11)$ MeV.
These results are shown in Fig.~\ref{fig:fD(s)}.
Fig.~\ref{fig:fD(s)} also shows the value of $f_D$ from
CLEO-c\cite{:2008sq}, and
 a recent preliminary HFAG world 
average\cite{hfagDs},
$f_{D_s}=263.9(6.7)$ MeV.
The results for theory and experiment are nicely compatible for
$f_D$ (as they are far the vast majority of lattice calculations),
but disagree significantly for $f_{D_s}$.
This  discrepancy is the largest discrepancy that has arisen
in lattice phenomenology.
It may be resolved by the lattice or experiment moving or changing their 
uncertainties.
The experimental average has come down in the last year, and the discrepancy was
3.8 sigma a year ago.
A recent CLEO-c number is lower still, quoting\cite{briere09}
$f_{D_s} = 259.5(7.3)$ MeV,
so it is quite possible that the discrepancy will simply disappear.
If theory and experiment were to remain inconsistent, 
the discrepancy could in principle
be an indication of new physics.\cite{Dobrescu:2008er}

\subsubsection{$D\rightarrow \pi l \nu , D\rightarrow K l \nu$ }
\begin{figure}
       \centerline{\includegraphics[width=.85\textwidth]{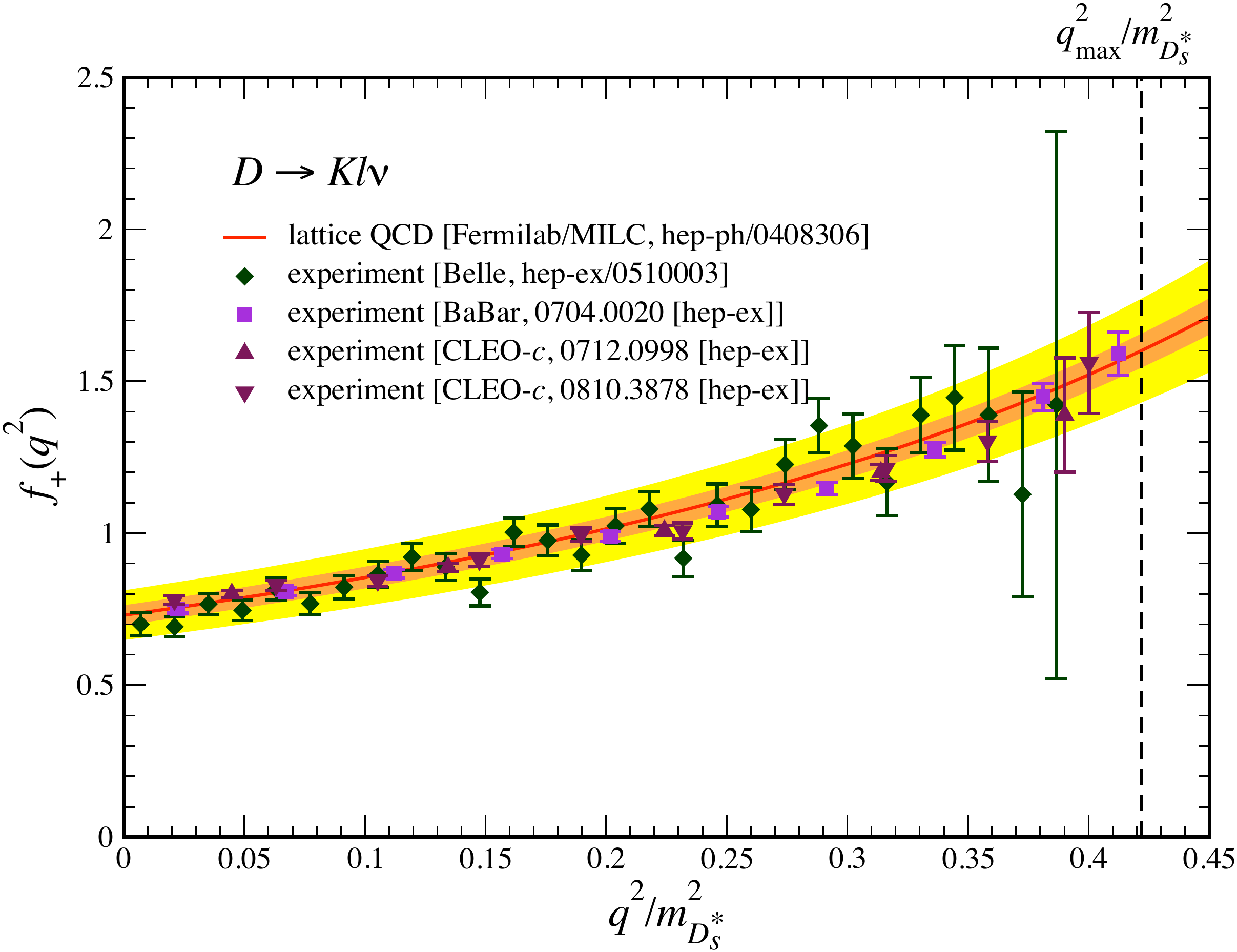}}
\caption{The shape of the form factor $f_+(q^2)$ predicted with lattice QCD
(orange and yellow bands), and as measured by Belle (green diamonds), 
by BaBar (magenta squares), and by CLEO (purple triangles).
}
\label{fig:DKshape}
\end{figure}

The shape of $D$ semileptonic decay was predicted by lattice 
calculations\cite{Aubin:2004ej,Okamoto:2004xg}
before its precise measurement, and subsequently confirmed by the
Focus, BaBar, and Belle experiments.
Fig.~\ref{fig:DKshape} shows the shape of the form factor $f_+(q^2)$ for the
decay $D\rightarrow K l \nu$ calculated on the 
lattice\cite{Aubin:2004ej,Okamoto:2004xg}
(orange and yellow bands), and as measured in experiment (points)
by Belle,\cite{Abe:2005sh} 
BaBar,\cite{Aubert:2007wg} and 
CLEO-c.\cite{:2007se,:2008yi}

The magnitude of the form factors  can be used to determine
$|V_{cd}|$ and $|V_{cs}|$.
At the conference Flavor Physics and CP Violation 2009, 
using CLEO-c data and the 2005 lattice 
results,\cite{Aubin:2004ej}
Xin obtained\cite{Xin:2009}
\begin{eqnarray}
  |V_{cs}|&=& 0.985(11)(103),   \\
  |V_{cd}|&=&   0.234 (7)(25),
\end{eqnarray}
where the errors are from experiment and theory.
Improved lattice calculations are underway for the theoretical errors.
Prospects for the theoretical uncertainty to be significantly reduced are
very good.

\subsection{$V_{cb}$ and $V_{ub}$
}
These can be directly determined from the leptonic and
exclusive semileptonic decays of $B$ mesons.

\subsubsection{$B\rightarrow D l \nu$ , $B\rightarrow D^* l \nu$ }

In analyses that assume the unitarity of the CKM matrix, 
$|V_{cb}|$ enters with a high power.
For example, the kaon mixing parameter $\epsilon$ is proportional to
$|V_{cb}|^4$, and $|V_{cb}|$ contributes as much uncertainty to the
theoretical prediction for $\epsilon$ as the  kaon mixing parameter $B_K$,
in spite of the fact that $|V_{cb}|$ is known to much higher precision.

Fortunately, it is possible to determine the form factors in
$B\rightarrow D l \nu$ and $B\rightarrow D^* l \nu$ decays much more
precisely than is possible in most leptonic and semileptonic decays.
The case of $B\rightarrow D l \nu$ is especially simple to analyze.
It is possible to connect $|V_{cb}|$ with the semileptonic form factors via
a quantity in which the uncertainties cancel almost completely in the
heavy-quark symmetry limit.
The form factor may be obtained from the double ratio:\cite{Hashimoto:1999yp}
\begin{equation}
|h(1)_+|^2 = 
\frac{<D|\overline{c}\gamma_0 b |\overline{B}>}
     {<D|\overline{c}\gamma_0 c |D>}
\frac{<\overline{B}|\overline{b}\gamma_0 c |D>}
     {<\overline{B}|\overline{b}\gamma_0 b |\overline{B}>}.
\end{equation}
The factors in the denominator are used in the vector current renormalization
required in this amplitude.
This ratio approaches one in the heavy-quark symmetry limit.
In lattice calculations,  uncertainties cancel almost completely in this limit.
For physical values of the $b$ and $c$ quarks, uncertainties are proportional
to deviations from the symmetry limit to high precision.
$h_+$ is the dominant term in the function ${\cal G}_{B\to D}$ for the decay
$B\rightarrow D l \nu$.
An unquenched lattice calculation gives
\cite{Okamoto:2004xg}
\begin{equation}
{\cal G}_{B\rightarrow D} (1)=1.074(18)(16).
\end{equation}
Using 
$|V_{cb}|{\cal G}(1) = (42.4\pm 1.6)\times 10^{-3}$
from HFAG,\cite{hfagichep08}
this produces
\begin{equation}
|V_{cb}| = \left(39.5 (1.5)_{\rm exp}(0.9)_{\rm theo} \right)\times 10^{-3},
\end{equation}
where the theory errors have been added in quadrature.  The total error for 
$|V_{cb}|$ from $B\rightarrow D l \nu$ is dominated by experiment.

Experimental errors for $B\rightarrow D^* l \nu$
are  smaller.
From the Review of Particle Physics,\cite{Amsler:2008zzb}
$|V_{cb}| {\cal F}(1) = \left(35.9\pm 0.8 \right) \times 10^{-3}$.
Hashimoto et al.\cite{Hashimoto:2001nb} defined a somewhat cumbersome 
combination of ratios from which ${\cal F}(1)$ can be determined.
Bernard et al.\cite{Bernard:2008dn}
investigated the much simpler ratio
\begin{equation}
|{\cal F}_{B\rightarrow D^*}(1)|^2 
= \frac{<D^*|\overline{c}\gamma_j\gamma_5 b|\overline{B}>}
{<D^*|\overline{c}\gamma_4 c|{D^*}>}
\frac{<\overline{B}|\overline{b}\gamma_j\gamma_5 c|D^*>}
{<\overline{B}|\overline{b}\gamma_4 b|\overline{B}>}
\end{equation}
from which ${\cal F}(1)$ can also be determined.
While calculational uncertainties do not cancel in this ratio as completely as
in the ratios introduced earlier,  they cancel to a high accuracy,
and because this quantity can be calculated ten to twenty times faster than the
set of ratios introduced earlier,\cite{Hashimoto:2001nb} 
determinations of ${\cal F}(1)$  from this ratio are ultimately more accurate.
Bernard et al. obtain
${\cal F}(1)=0.921(13)_{\rm stat}(20)_{\rm sys}.
$
Combining with the experimental result yields
\begin{equation}
|V_{cb}| = \left( 38.7  (0.9)_{\rm exp}  (1.0)_{\rm theo} \right) \times 10^{-3}.
\end{equation}

\subsubsection{$B\rightarrow \pi l \nu$ }

Comparison between theory and experiment for $B\rightarrow\pi\ell\nu$
has been more troublesome than for other lattice calculations in CKM physics.
Leptonic decays and $B\overline{B}$ mixing amplitudes are described by a 
single parameter.  
The semileptonic decays $B\rightarrow D^{(*)}\ell\nu$ and $K\rightarrow\pi\ell\nu$
can be described to high accuracy by a normalization and a slope.
For $B\rightarrow\pi\ell\nu$, on the other hand,  
the form factors have a complicated $q^2$ dependence.
Lattice data have covered only
the low momentum, high $q^2$ end of the pion momentum spectrum, and errors
are highly $q^2$ dependent and highly correlated in both theory and experiment.

It has long been understood that analyticity and unitarity can be
used to constrain the possible shapes of form factors.  
Consider mapping the variable $q^2$ onto a new variable, $z$, in the following way:
\begin{equation}
z(q^2, t_0) = \frac{\sqrt{1 - q^2/t_+}-\sqrt{1-t_0/t_+}}{\sqrt{1-q^2/t_+}+\sqrt{1-t_0/t_+}} ,
\label{eq:zee_var}
\end{equation}
where $t_+\equiv(m_B + m_\pi )^2$, $t_-\equiv(m_B - m_\pi )^2$, and $t_0$ is a 
free parameter.  Although this mapping appears complicated, it actually has a 
simple interpretation in terms of $q^2$;  this transformation maps $q^2 > t_+$ 
(the production region) onto $|z|=1$  and maps $q^2 < t_+$ (which includes the 
semileptonic region) onto real $z \in [-1,1]$.   
In the case of $B\rightarrow\pi\ell\nu$, the physical decay region is mapped
into roughly $-0.3<z<0.3$.
In terms of $z$, the form factors can be written in a simple form:
\begin{equation}
f(q^2) = \frac{1}{P(q^2) \phi(q^2,t_0)}  \sum_{k=0}^{\infty} a_k(t_0) z(q^2,t_0)^k.
\label{eq:z_exp}
\end{equation}
Most of the $q^2$ dependence is contained in the first two,
perturbatively calculable, factors.
 The Blaschke factor $P(q^2)$ is a function that contains subthreshold poles 
and the outer function $\phi(q^2,t_0)$ is an arbitrary analytic function (outside 
the cut from $ t_+ < q^2 < \infty$) which is chosen to give the
 series coefficients $a_k$ a simple form.
See\ \ \cite{Bailey:2008wp} \cite{Arnesen:2005ez}, and references therein for 
the explicit forms of these expressions.  With the proper choice of $\phi(q^2,t_0)$,
analyticity and unitarity require the $a_k$ to satisfy
\begin{equation}
\sum_{k=0}^{N} a_k^2 < \sim 1.
\label{eq:a_const}
\end{equation}
The fact that  $-0.3<z<0.3$ means that according to analyticity and unitarity,
only a few terms are required to describe the form factors to 1\%
accuracy.

Calculations have been performed  by the 
Fermilab Lattice and MILC collaborations\cite{Bailey:2008wp} 
using Fermilab  $b$ quarks, and by the HPQCD collaboration
using NRQCD $b$ quarks.\cite{Dalgic:2006dt}
\begin{figure}
\centerline{\includegraphics[width=.7\textwidth]{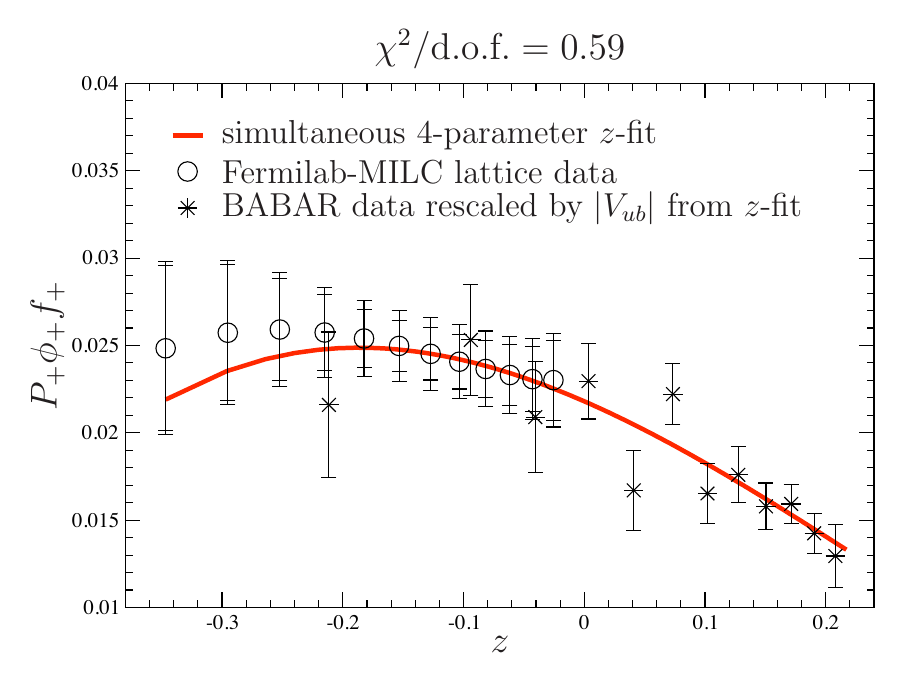}}
\caption{Results for the normalized
$B\rightarrow\pi\ell\nu$ form factor $P_+\phi_+f_+$ from the Fermilab/MILC
lattice calculations (circles) and BaBar (stars).
The solid red line is the results of a fully correlated simultaneous fit.
Requiring that lattice and experiment have the same normalization yields
$|V_{ub}|$.
}
\label{fig:Bpishape}
\end{figure}
Figure~\ref{fig:Bpishape} shows the result from Fermilab/MILC
of a fully correlated simultaneous $z$-fit
to the Fermilab/MILC lattice data and the BaBar 12-bin experimental 
results \cite{Aubert:2006px}, with $|V_{ub}|$ being a parameter in the fit.
%
Good fits may be obtained with just three terms in the expansion to the
lattice or the BaBar data separately, and to the combined
lattice and experimental data, with the result
\begin{equation}
\label{eq:answer}
	|V_{ub}| \times 10^3= 3.38 \pm 0.36 .
\end{equation}
%
Because the 10\% uncertainty comes from a simultaneous fit of the lattice and
experimental data, it contains  both the experimental and theoretical errors
in a way that is not simple to disentangle.
If we make the assumption that the error in 
$|V_{ub}|$ is dominated by the most 
precisely determined lattice point,
we can estimate that the contributions are roughly equally divided as 
$\sim$~6\% lattice statistical and chiral extrapolation (combined), 
$\sim$~6\% lattice systematic, and $\sim$~6\% experimental.
The largest lattice systematic uncertainties are heavy quark discretization,
the perturbative correction, and the uncertainty in $g_{B^*B\pi}$, all about 3\%.

HPQCD has obtained compatible results using NRQCD $b$ quarks 
and asqtad light quarks.\cite{Dalgic:2006dt}
They obtain consistent results from several types of fitting methods.
Applying their results to 2008 data from HFAG \cite{DiLodovico:2008uw} yields
\begin{eqnarray}
|V_{ub}| \times 10^3 & = & 3.40 \pm 0.20 ^{+0.59}_{-0.39},
\end{eqnarray}
where the first error is experimental and the second is from the lattice
calculations.

\subsubsection{$f_B$ and $f_{B_s}$}
$f_B$, and $f_{B_s}$ have been calculated by 
HPQCD\cite{Gamiz:2009ku,Wingate:2003gm,Gray:2005ad}
and by Fermilab/MILC.\cite{Bernard:2009wr,Bernard:2007zz}
HPQCD obtained $f_B=190(13)$ MeV and $f_{B_s}=231(15)$ MeV, 
and Fermilab/MILC obtained $f_B=195(11)$ MeV and $f_{B_s}=243(11)$ MeV.
It is complicated
 to know how to combine the uncertainties since the two calculations
share some uncertainties.
One could simply take the smaller of the two individual uncertainties 
as the combined uncertainty and use for an average
$f_B=193(11)$ MeV and $f_{B_s}=237(11)$ MeV.

The direct observation of the leptonic decay of the $B$ into a $\tau$ and its 
neutrino has become increasingly precise.
The world average\cite{Amsler:2008zzb}
 of $\Gamma(\tau^+\nu)/\Gamma_{\rm total}=1.4(4)\times10^{-4}$
allows a determination of $|V_{ub}|$ to 14\%.  While not (yet) competitive with the
determination for $B\rightarrow \pi l \nu$, this is a level that would have 
been hard to imagine a few years ago.

\subsection{$V_{ud}$ and $V_{us}$}
$|V_{ud}|$ and $|V_{us}|$ are directly related to the physics of pions and kaons.
The top row of the CKM matrix provides a precise test of the unitarity of 
the matrix via the relation
\begin{equation}
 |V_{ud}|^2+|V_{us}|^2+|V_{ub}|^2=1.
\label{firstUn}
\end{equation}
  Since $|V_{ub}|$ is around
$4\times 10^{-3}$, the third term's effects are negligible in this relation.
$|V_{ud}|$ can be obtained from pion leptonic decay and lattice calculations 
of the pion decay constant, but these determinations are not as accurate
as those from nuclear beta decays.
Lattice calculations of $f_\pi$ provide tests of lattice techniques.
From nuclear beta decay, we have
$|V_{ud}|=0.97425(22)$.\cite{Hardy:2008gy}
This contributes an uncertainty of $\pm0.00043$ to Eq.~(\ref{firstUn}).
Hence, a comparable level of uncertainty is of interest in $|V_{us}|$, an
uncertainty of around 0.5\%.  
Remarkably, because
$|V_{us}|$ may be determined from ratios in which many of the
theoretical uncertainties cancel 
and the remaining uncertainties are proportional to SU(3) flavor breaking,
this daunting level of accuracy is not out of reach.

$|V_{us}|$ may be obtained in two ways from ratios.
 $|V_{us}/V_{ud}|$ may be obtained via
the ratio of pion and kaon decay constants $f_\pi/f_K$.
$|V_{us}|$ may be directly obtained from a double ratio similar to that
used to obtain $|V_{cb}|$.  

\subsubsection{$f_\pi$ and $f_K$}
In 2004, Marciano emphasized the possibility of obtaining 
$|V_{us}/V_{ud}|$ from the ratio of the leptonic decay 
constants.\cite{Marciano:2004uf}
He provided a necessary radiative correction to the ratio and obtained
\begin{equation}
\frac{|V_{us}|}{|V_{ud}|}\frac{f_K}{f_\pi} = 0.2757(7).
\end{equation}

At Lattice 2008, kaon physics was reviewed by Lellouch.\cite{Lellouch:2009fg}
Most lattice uncertainties cancel in the ratio
 $f_K/f_\pi$.
The dominant uncertainty results from a combination of chiral extrapolation
and statistical error.
Lellouch performed a weighted average of the unquenched results and obtained
$f_K/f_\pi = 1.194(10)$.  When combined with Marciano's result,
this yields
\begin{equation}
\frac{|V_{us}|}{|V_{ud}|} = 0.2309(18),
\end{equation}
or about 0.8\% uncertainty, dominated by theory.



\subsubsection{$K\rightarrow \pi l \nu$}

Like $B\rightarrow D l \nu$,
the semileptonic form factor for the decay $K\rightarrow \pi l \nu$ may be
obtained from a double ratio.
In the case of $K\rightarrow \pi l \nu$,  flavor SU(3) 
symmetry ensures that most statistical and systematic errors
cancel in the symmetry limit, so that errors are very small in the physical case.
One starts from the double ratio
\begin{equation}
f_0(q^2_{\rm max}) = \frac{2\sqrt{M_K M_\pi}}{(M_K+M_\pi)}
  \frac{<\pi|V_0|K><K|V_0|\pi>}{<\pi|V_0|\pi><K|V_0|K>}.
\end{equation}
The denominator factors remove current renormalizations.
As shown by Becerivic et al.~\cite{Becirevic:2004ya},
by interpolating in $q^2$ to obtain $f_0(q^2=0)$ and using
$f_+(0)=f_0(0)$, one obtains the required form factor $f_+$ to high accuracy.
Lattice results for $f_+(0)$ were reviewed in  Lellouch\cite{Lellouch:2009fg},
with the average
$f_+(0) = 0.964(5)$.
The experimental result for Kl3 decay is\cite{BM}
$|V_{us}|\times f_+(0) = 0.21664(4)$.
This gives
\begin{equation}
|V_{us}| = 0.2247(12),
\end{equation}
or about 0.5\% accuracy.

Thus, the accuracies obtained for the unitarity test of Eqn.~(\ref{firstUn})
via these two methods
are reaching the required 0.5\% level.
Because the quantities involved are unusually simple lattice quantities, 
and because the necessary information can be obtained from ratios in which
most uncertainties cancel, the prospects are good for surpassing the required
accuracy soon.

In 1984, Leutwyler and Roos made the estimate
\cite{Leutwyler:1984je}
$f_+(0)=0.961 (8)$.
This estimate has stood the test of time remarkably well.
It is in agreement with modern lattice calculations,
and lattice calculations have only recently surpassed it in precision.
Some lattice calculations, such as heavy meson decay constants, have revealed
that QCD predicts something quite different from the old conventional 
wisdom that had been arrived at via quark models.
In $K\rightarrow \pi l \nu$ decay on the other hand,
Leutwyler and Roos's rough estimate of twenty years ago has proved
remarkably robust in the face of more fundamental calculations.

\subsection{$V_{td}$ and $V_{ts}$
}
$V_{td}$ and $V_{ts}$ cannot be determined from tree level processes,
but can be determined from loop effects in meson-antimeson mixing.
$B\overline{B}$ and $B_s\overline{B}_s$ mixing depend on 
$V_{td}$ and $V_{ts}$ in a simple way.
The kaon mixing parameter $\epsilon$ depends on a complicated mixture of 
quantities including $V_{td}$ and $V_{ts}$, and I will discuss it
 in this section, too.

\subsubsection{$B\overline{B}$ and $B_s\overline{B}_s$ mixing}
In $B_{(s)}$ mixing, the quantities $f_B\sqrt{B_B}$, $f_{B_s}\sqrt{B_{B_s}}$, and
$\xi\equiv f_{B_s}\sqrt{B_{B_s}}/(f_B\sqrt{B_B})$ are calculated.
The mixing amplitudes like$f_{B}\sqrt{B_{B}}$ are traditionally 
parametrized by a decay constant and
a ``bag'' parameter, but the amplitude is actually calculated as a single quantity.
Many of the usual lattice uncertainties cancel to a large extent in $\xi$.
Its uncertainty is dominated by a combination of chiral extrapolation and
statistics.
$f_{B_s}\sqrt{B_{B_s}}$ has only a mild chiral extrapolation (arising from the sea
 quarks), but it has a larger share of the usual lattice uncertainties, such
as discretization, operator matching, and lattice scale.
$\xi$ and $f_{B_s}\sqrt{B_{B_s}}$ therefore have relatively uncorrelated 
uncertainty budgets, and are the most suitable pair of quantities for inputs
into global fits.  $f_{B}\sqrt{B_{B}}$, by contrast, has a larger admixture
of all of these uncertainties.

The expressions for $B_{(s)}$ mixing can be written
\begin{equation}
\Delta M_{d(s)} \propto |V^*_{td(s)} V_{tb}|^2 f^2_{B_{d(s)}} \hat{B}_{B_{d(s)}},
\end{equation}
where $<\overline{B}^0_s|O^{s(d)}_L|B^0_s>= \frac{8}{3}M^2_{B_{s(d)}}f^2_{B_{s(d)}}B_{B_{s(d)}}(\mu),$ and \\
$O_L^{s(d)}=[\overline{b}^i\gamma_\mu(1-\gamma_5)s^i(d^i)]
[\overline{b}^j\gamma_\mu(1-\gamma_5)s^j(d^j)]$.
HPQCD have published results for the mixing amplitudes\cite{Gamiz:2009ku}
$f_{B_s}\sqrt{B_{B_s}} = 266(18)$ MeV and
$f_{B}\sqrt{B_{B}} = 216(15)$ MeV.

The most precise constraint on the $\overline{\rho}$-$\overline{\eta}$ plane
from $B$ and $B_{s}$ mixing
can be obtained from the combination
\begin{equation}
\frac{|V_{td}|}{|V_{ts}|} =
  \xi \sqrt{\frac{\Delta M_d}{\Delta M_s}  \frac{M_{B_s}}{M_{B_d}}}.
\label{Vtdts}
\end{equation}
For $\xi$, HPQCD has\cite{Gamiz:2009ku}
$\xi = 1.258(33)$.
A preliminary result from Fermilab/MILC gives
$\xi = 1.205(50)$.\cite{ToddEvans:2008}  
The uncertainty in this second number is expected to drop in the final result
expected later this year.
Combining the HPQCD result with Eq.~(\ref{Vtdts}) gives
\begin{equation}
\frac{|V_{td}|}{|V_{ts}|} = 0.214(1)(5),
\end{equation}
where the first error is experimental and the second is theoretical.
In spite of the high precision already achieved theoretically for $\xi$,
the even higher experimental precision 
for $\Delta M_d/\Delta M_s$
means that a much larger
payoff in constraining the CKM matrix will result from further improved
lattice calculations, which should be possible.

One of the most important possible outcomes of high precision CKM constraints
would be an inconsistency in the constraints, indicating new physics.
The fact that $B$ and $B_s$ mixing originate in loop effects has led to
speculation that they may be more sensitive to new physics in the loops than
are tree-level processes.
In $B\overline{B}$ mixing, 
there are five operators in the most general basis for four-quark
scattering  that can arise in BSM physics.
These have been tabulated and evaluated in the quenched approximation by
Becirevic et al.\cite{Becirevic:2002qr} 
It is straightforward to repeat these calculations in unquenched calculations,
and efforts to do so are underway.

\subsubsection{A comment on $D\overline{D}$ mixing on the lattice}
The calculation of the short-distance part of $D\overline{D}$ 
mixing is identical to the $B\overline{B}$ mixing case just discussed.
These calculations have received less attention on the lattice than
the analogous calculations in the $B$ system because
in $D\overline{D}$ mixing, there are long-distance contributions
from intermediate states.
To calculate these on the lattice in a naive way would require a
four-point function, as opposed to the three-point functions used in current
$B\overline{B}$ mixing calculations.
This would require a number of numerical operations proportional to the 
lattice volume, $V$, squared, as opposed to the ${\cal O}(V)$ operations used
in current calculations.  This is beyond the capability of current computers.
It is probably possible to attack long-distance effects with 
more intelligent methods,
but these have not been worked out.

\subsubsection{$K\overline{K}$ mixing}
The formula for the $K\overline{K}$ mixing parameter $\epsilon$ is
\begin{eqnarray}
|\epsilon|&=&\frac{G^2_Ff_K^2m_Km_W^2}{12\sqrt{2}\pi^2\Delta m_K}
   \hat{B}_K  [\eta_cS(x_c){\rm Im}[(V_{cs}V^*_{cd})^2]       \nonumber\\  
          &+&\eta_t S(x_t){\rm Im}[(V_{ts}V^*_{td})^2]
          +2 \eta_{ct} S(x_c,x_t){\rm Im}[(V_{cs}V^*_{cd}V_{ts}V^*_{td})]   ] .
\end{eqnarray}
This depends on the renormalization group invariant bag parameter $\hat{B}_K$ and on the
CKM matrix elements $V_{cs}$, $V_{cd}$, $V_{ts}$, and $V_{td}$ and on several
other correction factors with uncertainties.
Many of these uncertainties contribute to the 
uncertainty in the bound that kaon mixing can
give in the $\rho$-$\eta$ plane, which makes the prospects for improving the bound
more problematic to estimate.

World results for $\hat{B}_K$ were reviewed by Lellouch at
Lattice 2009, who obtained\cite{Lellouch:2009fg,Aubin:2009jh}
$\hat{B}_K = 0.723(37)$.
The largest uncertainties in constraining the $\overline{\rho}$-$\overline{\eta}$ 
from $\epsilon$ arise from $\hat{B}_K$ and $V_{cb}$.\cite{Amsler:2008zzb}
(The latter is in the sense that, assuming CKM unitarity,
$(V_{ts}V^*_{td})^2$ is proportional to the
Wolfenstein parameter $A^4$, which in turn is proportional to  $|V_{cb}|^4$.)
However, as illustrated in Lellouch,\cite{Lellouch:2009fg}
even if these uncertainties were to be reduced to zero in expression 
for $\epsilon$, there would still be significant uncertainty  from the rest of 
the parameters.
In Fig.~\ref{fig:rhoeta}, one sees that the constraint in
the $\overline{\rho}$-$\overline{\eta}$ plane from $\epsilon$ is less stringent
that those from $B$ and $B_s$ mixing and from $V_{ub}$ inclusive and exclusive.
Improving this bound will be possible, but complicated.

\subsection{$V_{tb}$
}
This is the only element of the
 CKM Matrix about which lattice QCD has nothing to say.

\section{Present and  future lattice calculations}

\subsection{The current generation of lattice QCD calculations}
The current generation of lattice QCD calculations serves both to
demonstrate the correctness of lattice calculations and to deliver
quantities for physics that can only be obtained with lattice QCD.
Foundations are also currently being laid for calculations
in strongly interacting Beyond-the-Standard-Model theories, should they
be discovered.
One obvious goal of the current generation of lattice calculations
is to demonstrate the correctness of lattice methods, such as by
correctly post-dicting  the hadron spectrum.
Some of the weak matrix elements I have discussed can also serve to do 
this by constructing CKM-independent ratios of physical quantities.
Occasionally, it has been possible to make predictions, even in low energy
hadronic physics, such as in the shape of the form factor in
$D_s\rightarrow K l \nu$ decay that I discussed.
Also of greater interest to the general community may be the resolutions of the
contradictions that have  occurred in a few places between lattice results 
and other results.  
 The lattice results for $f_{D_s}$ lie significantly below the experimental
results.  There are also interesting two-sigma tensions between inclusive
and exclusive results for $|V_{ub}|$ and $|V_{cb}|$.
It is not known at present how these will play out.

The lattice determinations of CKM matrix elements that are the topic
of this article are among
 the most important current deliverables of lattice gauge theory
for particle physics and for the world of physics in general.
Some of the the fundamental parameters of the Standard Model 
can be directly determined only with lattice QCD, 
including the light quark masses and the constraints on the CKM
matrix from $K\overline{K}$, $B\overline{B}$, and $B_s\overline{B}_s$
mixing. 


Some flavor issues may have their resolution in Beyond-the-Standard-Model
physics.  Lattice calculations are beginning to address these.
One example that has become tractable with 
the current  technology of lattice QCD is the
search for near-conformal behavior in gauge theories.\cite{Fleming:2008gy}
One explanation for the smallness of flavor-changing neutral currents
in the context of Technicolor models of dynamical breaking of the weak gauge 
symmetry involves an almost conformal fixed point in the symmetry breaking
sector.  Lattice calculations have begun to search for such theories.

\subsection{The coming generation of lattice calculations.}

I have not said much about prospects for future improvements in the results
I have discussed.  This is because the prospects for almost all of them are 
excellent.
Most of the most important lattice results in CKM physics involve hadronically
stable mesons.
These are the simplest hadrons to study on the lattice (or elsewhere), and
there are no known impediments to higher precision.
Improvements to methods, algorithms, and computers are putting errors
such as discretization and chiral extrapolation under increasingly good 
control.
Uncertainties due to operator normalizations fall more slowly with brute
force computer power, but here also we are in good shape.
Lattice perturbation theory converges more or less as well does dimensionally
regulated perturbative QCD.\cite{Lepage:1992xa}
On the lattice, however, there are also nonperturbative renormalization methods
available that are not available in the continuum,\cite{Luscher:1998pe}
putting short-distance QCD under better control on the lattice than in the 
continuum.
Achieving the sub-per cent precision required by experiment in 
$K\overline{K}$, $B\overline{B}$, and $B_s\overline{B}_s$ mixing for bounding
the $\overline{\rho}$-$\overline{\eta}$ plane is by no means an unreachable goal. 

An increasing variety of physical quantities will become tractable
in the coming years.
I have focused in this article on processes with single mesons in the final
state, because these are the most precise lattice CKM calculations.
Exclusive processes with more than one hadron are more complicated because
the translation between Euclidean final states and Minkowskian final states
are more complicated than for single-hadron states,
but methods exist for some of these.
First-principles methods have been worked out for the two-pion decays of 
$K$ mesons, which are needed to extract CKM properties from
$\epsilon'/\epsilon$\cite{Lellouch:2000pv}.  
Chiral perturbation theory can be used to help analyze 
final states, since the decay pions have relatively small momentum.
These calculations are somewhat more cumbersome than single hadron
calculations and will require more computation,
 but can ultimately be done with current methods,
and prospects for progress are good.
For interesting decays like $B\rightarrow \pi\pi$, on the other hand,
no practical methods exist, and new methods must be developed.
The long-distance contributions to $D\overline{D}$ mixing have not yet
been addressed with lattice calculations.
The well-advanced topics I discussed in the previous section are done with
two- and three-point correlation functions.
The higher-point  correlation functions required
in $D\overline{D}$ mixing and similarly in the ``light-by-light'' 
scattering nonperturbative contributions to $g-2$ for the muon
require more advanced methods.
Those for light-by-light scattering are now being worked out.

Some Beyond-the-Standard-Model strongly coupled gauge theories are within
the reach of current methods, as already discussed.\cite{Fleming:2008gy}
Others, such as the simplest strongly coupled supersymmetric gauge theories
are tractable with straightforward extensions of current 
methods and are being actively investigated.\cite{Giedt:2009yd}
Others will require more significant improvements to methods.
These are all being given increased attention as the LHC era dawns.
How new data from the LHC will affect these developments is impossible to foresee.
With increasingly precise results and a broader variety of applications, 
 lattice gauge theory will continue to
 play an expanding role in particle physics in the LHC era.

\section*{Acknowledgements}
I would like to thank Elvira Gamiz, Jack Laiho,
 and Ruth Van de Water for helpful comments.

%


\begin{thebibliography}{99}


\bibitem{Kobayashi:1973fv}
  M.~Kobayashi and T.~Maskawa,
  ``CP Violation In The Renormalizable Theory Of Weak Interaction,''
  Prog.\ Theor.\ Phys.\  {\bf 49}, 652 (1973).


\bibitem{Cabibbo:1963yz}
  N.~Cabibbo,
  ``Unitary Symmetry and Leptonic Decays,''
  Phys.\ Rev.\ Lett.\  {\bf 10}, 531 (1963).

\bibitem{Wolfenstein:1983yz}
  L.~Wolfenstein,
  ``Parametrization Of The Kobayashi-Maskawa Matrix,''
  Phys.\ Rev.\ Lett.\  {\bf 51}, 1945 (1983).

\bibitem{ckmfitter} CKMfitter Group (J. Charles et al.), 
Eur. Phys. J. C41, 1-131 (2005) [hep-ph/0406184],
updated March, 2009, http://ckmfitter.in2p3.fr.

\bibitem{Lellouch:2009fg}
  L.~Lellouch,
  ``Kaon physics: a lattice perspective,''
  arXiv:0902.4545 [hep-lat].


\bibitem{Gamiz:2008iv}
  E.~Gamiz,
  ``Heavy flavour phenomenology from lattice QCD,''
  arXiv:0811.4146 [hep-lat].

\bibitem{Follana:2007uv}
  E.~Follana, C.~T.~H.~Davies, G.~P.~Lepage and J.~Shigemitsu  [HPQCD
                 Collaboration and UKQCD Collaboration],
  ``High Precision determination of the pi, K, D and Ds decay constants   from
  lattice QCD,''
  Phys.\ Rev.\ Lett.\  {\bf 100}, 062002 (2008)
  [arXiv:0706.1726 [hep-lat]].


\bibitem{Follana:2006rc}
  E.~Follana {\it et al.}  [HPQCD Collaboration and UKQCD Collaboration],
  ``Highly Improved Staggered Quarks on the Lattice, with Applications to Charm
  Physics,''
  Phys.\ Rev.\  D {\bf 75} (2007) 054502
  [arXiv:hep-lat/0610092].

\bibitem{Blossier:2009bx}
  B.~Blossier {\it et al.},
  ``Pseudoscalar decay constants of kaon and D-mesons from Nf=2 twisted mass
  Lattice QCD,''
  arXiv:0904.0954 [hep-lat].

\bibitem{Bernard:2009wr}
  C.~Bernard {\it et al.},
  ``B and D Meson Decay Constants,''
  PoS {\bf LATTICE2008}, 278 (2008)
  [arXiv:0904.1895 [hep-lat]].


\bibitem{Bernard:2007zz}
  C.~Bernard {\it et al.}  [Fermilab Lattice, MILC and HPQCD Collaborations],
  ``The decay constants $f_B$ and $f_{D^+}$ from three-flavor lattice QCD,''
  PoS {\bf LAT2007}, 370 (2007).

\bibitem{:2008sq}
  B.~I.~Eisenstein {\it et al.}  [CLEO Collaboration],
  ``Precision Measurement of $B(D^+ \rightarrow \mu^+ \nu)$ and the Pseudoscalar Decay
  Constant $f_{D^+}$,''
  Phys.\ Rev.\  D {\bf 78}, 052003 (2008)
  [arXiv:0806.2112 [hep-ex]].

\bibitem{hfagDs} Talk of A.~Schwartz,
``$B$ and $D_s$ Decay Constants 
  from Belle and Babar'', at 
10th Conference on the Intersections  
of Particle and Nuclear Physics, 
Torrey Pines, San Diego, 
May 29, 2009.

\bibitem{briere09} Talk of Roy Briere, ``Leptonic D Decays'', at
  Flavor Physics and CP Violation 09, May, 2009.

\bibitem{Dobrescu:2008er}
  B.~A.~Dobrescu and A.~S.~Kronfeld,
  ``Accumulating evidence for nonstandard leptonic decays of $D_s$ mesons,''
  Phys.\ Rev.\ Lett.\  {\bf 100}, 241802 (2008)
  [arXiv:0803.0512 [hep-ph]].

\bibitem{Aubin:2004ej}
  C.~Aubin {\it et al.}  [Fermilab Lattice Collaboration and MILC
                  Collaboration and HPQCD Collab],
  ``Semileptonic decays of D mesons in three-flavor lattice QCD,''
  Phys.\ Rev.\ Lett.\  {\bf 94}, 011601 (2005)
  [arXiv:hep-ph/0408306].

\bibitem{Okamoto:2004xg}
  M.~Okamoto {\it et al.},
  ``Semileptonic $D\rightarrow \pi / K$ and $B \rightarrow \pi / D$ decays in 2+1 flavor lattice
  QCD,''
  Nucl.\ Phys.\ Proc.\ Suppl.\  {\bf 140}, 461 (2005)
  [arXiv:hep-lat/0409116].


\bibitem{Abe:2005sh}
  K.~Abe {\it et al.}  [BELLE Collaboration],
  ``Measurement of $D^0 \rightarrow \pi l \nu (K l \nu)$ and their form factors,''
  arXiv:hep-ex/0510003.


\bibitem{Aubert:2007wg}
  B.~Aubert {\it et al.}  [BABAR Collaboration],
  ``Measurement of the hadronic form-factor in $D^0 \to K^{-} e^{+} \nu_{e}$
  1,''
  arXiv:0704.0020 [hep-ex].

\bibitem{:2007se}
  D.~Cronin-Hennessy {\it et al.}  [CLEO Collaboration],
  ``A Study of the Decays 
  $D^0\rightarrow \pi^- e^+ \nu_e$, $D^0\rightarrow K^- e^+ \nu_e$, 
  $D^+ \rightarrow  \pi^0 e^+ \nu_e$, and 
  $D^+ \rightarrow \overline{K}^0 e^+ \nu_e$,''
  Phys.\ Rev.\ Lett.\  {\bf 100}, 251802 (2008)
  [arXiv:0712.0998 [hep-ex]].

\bibitem{:2008yi}
  J.~Y.~Ge {\it et al.}  [CLEO Collaboration],
  ``Study of $D^0\rightarrow \pi^- e^+ \nu_e$, 
  $D^+ \rightarrow \pi^0 e^+ \nu_e$, 
  $D^0 \rightarrow K^- e^+ \nu_e$, and
  $D^+ \rightarrow \overline{K}^0 e^+ \nu_e$
  in Tagged Decays of the psi(3770) Resonance,''
  arXiv:0810.3878 [hep-ex].

\bibitem{Xin:2009}
  Talk by Bo Xin, ``Charm Semileptonic Decays'', at Flavor Physics and CP
  Violation, 2009''.
  After this article was submitted, the results referred to in the text were
  published in the 500th paper of the CLEO Collaboration,
  ~D.~Besson  [The CLEO Collaboration],
  ``Improved measurements of D meson semileptonic decays to pi and K mesons,''
  arXiv:0906.2983 [hep-ex].

\bibitem{Hashimoto:1999yp}
  S.~Hashimoto, A.~X.~El-Khadra, A.~S.~Kronfeld, P.~B.~Mackenzie, S.~M.~Ryan and J.~N.~Simone,
  ``Lattice {QCD} calculation of $\overline{B} \rightarrow D l \overline{\nu}$
 decay form factors  at zero recoil,''
  Phys.\ Rev.\  D {\bf 61}, 014502 (1999)
  [arXiv:hep-ph/9906376].

\bibitem{hfagichep08} HFAG, ICHEP08 update, \\
  http://www.slac.stanford.edu/xorg/hfag/semi/ichep08/home.shtml.

\bibitem{Amsler:2008zzb}
  C.~Amsler {\it et al.}  [Particle Data Group],
  ``Review of particle physics'' (RPP),
  Phys.\ Lett.\  B {\bf 667}, 1 (2008).

\bibitem{Hashimoto:2001nb}
  S.~Hashimoto, A.~S.~Kronfeld, P.~B.~Mackenzie, S.~M.~Ryan and J.~N.~Simone,
  ``Lattice calculation of the zero recoil form factor of 
  $\overline{B} \rightarrow D^* l \overline{\nu}$: 
  Toward a model independent determination of $|V(cb)|$,''
  Phys.\ Rev.\  D {\bf 66}, 014503 (2002)
  [arXiv:hep-ph/0110253].

\bibitem{Bernard:2008dn}
  C.~Bernard {\it et al.},
  ``The $\bar{B} \to D^{*} \ell \bar{\nu}$ form factor at zero recoil from
  three-flavor lattice QCD: A Model independent determination of $|V_{cb}|$,''
  Phys.\ Rev.\  D {\bf 78}, 094505 (2008)
  [Phys.\ Rev.\  D {\bf 79}, 014506 (2009)]
  [arXiv:0808.2519 [hep-lat]].

\bibitem{Bailey:2008wp}
  J.~A.~Bailey {\it et al.},
  ``The $B \to \pi \ell \nu$ semileptonic form factor from three-flavor lattice
  QCD: A Model-independent determination of $|V_{ub}|$,''
  arXiv:0811.3640 [hep-lat].

\bibitem{Arnesen:2005ez}
  M.~C.~Arnesen, B.~Grinstein, I.~Z.~Rothstein and I.~W.~Stewart,
  ``A precision model independent determination of |V(ub)| from B --> pi e
  nu,''
  Phys.\ Rev.\ Lett.\  {\bf 95}, 071802 (2005)
  [arXiv:hep-ph/0504209].

\bibitem{Dalgic:2006dt}
  E.~Dalgic, A.~Gray, M.~Wingate, C.~T.~H.~Davies, G.~P.~Lepage and J.~Shigemitsu,
  Phys.\ Rev.\  D {\bf 73}, 074502 (2006)
  [Erratum-ibid.\  D {\bf 75}, 119906 (2007)]
  [arXiv:hep-lat/0601021].

\bibitem{Aubert:2006px}
  B.~Aubert {\it et al.}  [BABAR Collaboration],
  ``Measurement of the $B^0 \to \pi^{-} \ell^{+} \nu$ form-factor shape and
  branching fraction, and determination of $|V_{ub}|$ with a loose neutrino
  reconstruction technique,''
  Phys.\ Rev.\ Lett.\  {\bf 98}, 091801 (2007)
  [arXiv:hep-ex/0612020].






\bibitem{DiLodovico:2008uw}
  F.~Di Lodovico,
  ``A Review of the Magnitudes of the CKM Matrix Elements,''
  Int.\ J.\ Mod.\ Phys.\  A {\bf 23}, 4945 (2008)
  [arXiv:0811.3540 [hep-ex]].

\bibitem{Gamiz:2009ku}
  E.~Gamiz, C.~T.~H.~Davies, G.~P.~Lepage, J.~Shigemitsu and M.~Wingate
                  [HPQCD Collaboration],
  ``Neutral $B$ Meson Mixing in Unquenched Lattice QCD,''
  arXiv:0902.1815 [hep-lat].

\bibitem{Gray:2005ad}
  A.~Gray {\it et al.}  [HPQCD Collaboration],
  ``The B Meson Decay Constant from Unquenched Lattice QCD,''
  Phys.\ Rev.\ Lett.\  {\bf 95}, 212001 (2005)
  [arXiv:hep-lat/0507015].

\bibitem{Wingate:2003gm}
  M.~Wingate, C.~T.~H.~Davies, A.~Gray, G.~P.~Lepage and J.~Shigemitsu,
  ``The B/s and D/s decay constants in 3 flavor lattice QCD,''
  Phys.\ Rev.\ Lett.\  {\bf 92}, 162001 (2004)
  [arXiv:hep-ph/0311130].

\bibitem{Hardy:2008gy}
  J.~C.~Hardy and I.~S.~Towner,
  ``Superallowed 0+ to 0+ nuclear beta decays: A new survey with precision
  tests of the conserved vector current hypothesis and the standard model,''
  arXiv:0812.1202 [nucl-ex].

\bibitem{Marciano:2004uf}
  W.~J.~Marciano,
  ``Precise determination of $|V(us)|$ from lattice calculations of  pseudoscalar
  decay constants,''
  Phys.\ Rev.\ Lett.\  {\bf 93}, 231803 (2004)
  [arXiv:hep-ph/0402299].


\bibitem{Becirevic:2004ya}
  D.~Becirevic {\it et al.},
  ``The $K \rightarrow \pi$ vector form factor at zero momentum transfer on the
  lattice,''
  Nucl.\ Phys.\  B {\bf 705}, 339 (2005)
  [arXiv:hep-ph/0403217].

\bibitem{BM} E.~Blucher and W.~J.~Marciano, in RPP\cite{Amsler:2008zzb}.

\bibitem{Leutwyler:1984je}
  H.~Leutwyler and M.~Roos,
  ``Determination Of The Elements V(Us) And V(Ud) Of The Kobayashi-Maskawa
  Matrix,''
  Z.\ Phys.\  C {\bf 25}, 91 (1984).


\bibitem{ToddEvans:2008}
  R.~Todd Evans, A.~X.~El-Khadra and E.~Gamiz, 
  PoS {\bf LAT2008}, 52 (2008)

\bibitem{Becirevic:2002qr}
  D.~Becirevic, V.~Gimenez, V.~Lubicz, G.~Martinelli, M.~Papinutto and J.~Reyes
                  [SPQcdR collaboration],
  ``Non-perturbative renormalization of four fermion operators and B0 anti-B0
  mixing with Wilson fermions,''
  Nucl.\ Phys.\ Proc.\ Suppl.\  {\bf 119}, 619 (2003)
  [arXiv:hep-lat/0209131].




\bibitem{Aubin:2009jh} For a more recent high-precision result, see
  C.~Aubin, J.~Laiho and R.~S.~Van de Water,
  ``The neutral kaon mixing parameter $B_K$ from unquenched mixed-action lattice
  QCD,''
  arXiv:0905.3947 [hep-lat].

\bibitem{Fleming:2008gy}
  For a review of dynamical electro-weak
 symmetry breaking on the lattice, see G.~T.~Fleming,
  ``Strong Interactions for the LHC,''
  PoS {\bf LATTICE2008}, 021 (2008)
  [arXiv:0812.2035 [hep-lat]].

\bibitem{Lepage:1992xa}
  G.~P.~Lepage and P.~B.~Mackenzie,
  ``On The Viability Of Lattice Perturbation Theory,''
  Phys.\ Rev.\  D {\bf 48}, 2250 (1993)
  [arXiv:hep-lat/9209022].

\bibitem{Luscher:1998pe}
  See M.~Luscher,
  ``Advanced lattice QCD,''
  arXiv:hep-lat/9802029, and references therein.

\bibitem{Lellouch:2000pv}
  L.~Lellouch and M.~Luscher,
  ``Weak transition matrix elements from finite-volume correlation
  functions,''
  Commun.\ Math.\ Phys.\  {\bf 219}, 31 (2001)
  [arXiv:hep-lat/0003023].


\bibitem{Giedt:2009yd}
  For a review of lattice supersymmetry, see J.~Giedt,
  ``Progress in four-dimensional lattice supersymmetry,''
  arXiv:0903.2443 [hep-lat].

\end{thebibliography}
\end{document}